\documentclass[twocolumn,nofootinbib]{revtex4-1}
\usepackage{latexsym,epsfig,amssymb, amsmath}

\newcommand{\cN}{{\cal N}}

\def\bfone{\relax{\rm 1\kern-.35em 1}}

\newcommand{\be}{\begin{equation}}
\newcommand{\ee}{\end{equation}}
\newcommand{\ben}{\begin{displaymath}}
\newcommand{\een}{\end{displaymath}}
\newcommand{\bea}{\begin{eqnarray}}
\newcommand{\eea}{\end{eqnarray}}

\newcommand{\bean}{\begin{eqnarray*}}
\newcommand{\eean}{\end{eqnarray*}}

\makeatletter \@addtoreset{equation}{section} \makeatother

\begin{document}

\title{Flux Compactifications, Gauge Algebras and De Sitter}

\author{Giuseppe Dibitetto$^1$, Rom\'{a}n Linares$^2$ and Diederik Roest$^1$}

\affiliation{~ \\ $^1$Centre for Theoretical Physics,\\
University of Groningen, \\
Nijenborgh 4, 9747 AG Groningen, The Netherlands\\
{\small {\{ g.dibitetto, d.roest \} @rug.nl}} \\
$^2$Departamento de F\'{\i}sica, \\
Universidad Aut\'onoma Metropolitana Iztapalapa,\\
San Rafael Atlixco 186, C.P. 09340, M\'exico D.F., M\'exico\\
{\small {lirr@xanum.uam.mx}}}

\begin{abstract}
The introduction of (non-)geometric fluxes allows for $\cN = 1$ moduli
stabilisation in a De Sitter vacuum. The aim of this letter is to assess to what extent this
is true in $\cN = 4$ compactifications. First we identify
the correct gauge algebra in terms of gauge and (non-)geometric fluxes.
We then show that this algebra does not lead to any of the known gaugings with De Sitter solutions. In particular,
the gaugings that one obtains from flux compactifications involve non-semi-simple algebras, while the known gaugings with
De Sitter solutions consist of direct products of (semi-)simple algebras.
\end{abstract}

\maketitle

\section{Introduction}

Over the years superstring compactifications have been
investigated from many different perspectives. The possibility of
including various types of fluxes allows for different effective
descriptions of four-dimensional physics
(see e.g.~\cite{Grana, Douglas, Blumenhagen}).
The effective theories that are in general obtained by flux
compactifications in string theory are gauged supergravities. The
residual amount of supersymmetries that these four-dimensional
theories have depends on the internal manifold chosen for the
compactification and on the presence of local sources such as
orientifold planes, branes and Kaluza-Klein monopoles. A natural

question in this context is whether a vacuum state with positive
cosmological constant and spontaneously broken supersymmetry can
possibly arise. This would be relevant in order to embed some
cosmological features of our four-dimensional physics inside string
theory, such as slow-roll inflation and late-time acceleration of
universe.

In the context of type IIA string theory a number of no-go results
\cite{Hertzberg:2007wc, Silverstein, Haque:2008jz, Caviezel:2008tf,
Flauger:2008ad, Danielsson, Zagermann} essentially forbid the
existence of De Sitter vacua as long as a limited list of fluxes is
considered. Some of these results have been obtained in the case of
$SU(3)$ structure manifolds. Further recent works have investigated
the link between $\cN=4$ gauged supergravity and string theory
background fluxes in the presence of orientifold planes
\cite{Roest:2009dq, Dall'Agata:2009gv}. Such an analysis shows that
the $\cN=4$ supergravity side allows for much more freedom at the
level of deformations of the theory with respect to what is actually
possible in purely geometric backgrounds of string theory. In other
words, given a certain $\cN=4$ gauging, it is a highly non-trivial
question whether such a gauging has a higher dimensional origin in
terms of purely geometric and gauge fluxes. Due to this, the
so-called non-geometric fluxes \cite{STW, Aldazabal:2006up} (and, relatedly, doubled geometry \cite{Hull1,Hull2}) 
have been introduced in the literature as flux parameters which are T-
and S-dual to the known ones. This basically arises from the concept
of mirror symmetry as a way of extending dualities in the presence
of fluxes \cite{Grana:2006hr}. Using non-geometric fluxes, full
stabilisation of all moduli has been achieved in De Sitter vacua in
an $\cN = 1$ context \cite{deCarlos1,deCarlos2}.

In the present letter we first review the gauge algebra of $\cN=4$
gauged supergravity and its formulation in terms of the embedding
tensor (section 2). Secondly, we come to the identification of the
correct $\cN = 4$ gauge algebra in terms of fluxes (section 3). Even
making use of non-geometric fluxes, one cannot access any of the
gaugings of $\cN=4$ supergravity that are known to give rise to De
Sitter solutions \cite{deRoo:2002jf, deRoo:2003rm} (section 4). This
means that these gaugings do not have a higher dimensional origin
and cannot be understood in terms of a string theory background, not
even a non-geometric one. The argument shown later in this letter is
very simple and is obtained in the IIB duality frame with O3-planes;
this is a very convenient one because only four types of fluxes are
allowed by the orientifold projection, including non-geometric
fluxes. What we show is that the flux-induced gauge algebra is
always non-semi-simple due to the presence of an Abelian ideal. None
of the known examples of gaugings giving rise to De Sitter solutions
fall in this class of flux-induced algebras. In the conclusions we
suggest a possibility how one could evade this no-go theorem
(section 5).

\section{Gauge algebras in $\cN = 4$}

Half-maximal $\cN = 4$ supergravity corresponds to the low-energy
effective description of e.g.~ten-dimensional type I string theory
on a torus, or of type II string theories on $T^2 \times K3$ or on a
torus in the presence of an orientifold plane. The theory consists
of a supergravity multiplet and an additional number of vector
multiplets, which for our purposes will be six. In this case the
theory enjoys a global symmetry
 \begin{align}
  SL(2) \times SO(6,6) \,. \label{globsym}
 \end{align}
The doublet representation of $SL(2)$ will be denoted by $\alpha$,
whereas the fundamental representation of $SO(6,6)$ will be given by
$M$. We will take the corresponding metric to be
 \begin{align}
  \eta_{MN} =  \left(
   \begin{array}{cc} & \mathbb{I}_6 \\ \mathbb{I}_6 & \end{array} \right) \,, \quad M=(1,\ldots,6,\bar
1,\ldots, \bar 6) \,, \label{eta}
 \end{align}
i.e.~we use light-cone coordinates.

The bosonic fields form representations of this global symmetry
group. The scalars form a coset manifold based on \eqref{globsym}
and hence split up in two parts, of dimensions 2 and 36,
respectively. The vectors $A_{M \alpha}$ transform in the
fundamental representation of $SO(6,6)$. Furthermore, a crucial
point is that the electric and magnetic parts of the vectors
transform as doublets of $SL(2)$.

In the ungauged theory, there are Abelian gauge transformations
associated with every gauge vector. In other words, the theory has a
$U(1)^{12}$ gauge symmetry, in addition to the global symmetry
\eqref{globsym}. If one wants to include the magnetic part of the
vectors as well, one could even say that the theory has a
$U(1)^{24}$ gauge symmetry. However, this is only a symmetry of the
equations of motion, as the Lagrangian is formulated in terms of the
electric gauge potentials only.

The only deformations of this theory are the gaugings of some
subgroup of the global symmetry group \eqref{globsym}. These are
parametrised by the components of the so-called embedding tensor
\cite{Samtleben}. For $\cN = 4$ these components consist of
$\xi_{\alpha M}$ and $f_{\alpha MNP}$ \cite{Schon:2006kz}, where the
latter is completely antisymmetric in its $SO(6,6)$ indices. We will
restrict to the case with $\xi_{\alpha M} = 0$, implying that the
gauge group is restricted to act within $SO(6,6)$. In this case the
commutation relations read
  \begin{align}
   [ X^{M \alpha}, X^{N \beta} ] = f^{\alpha MN}{}_P X^{P \beta} \,, \label{comm}
  \end{align}
where $X^{M \alpha}$ is the generator corresponding to the gauge vector $A_{M\alpha}$.

The deformation parameters need to satisfy certain consistency
constraints which are called quadratic constraints. One way to
derive these is by requiring the embedding tensor components to be
invariant under gauge transformation. This results in
\cite{Schon:2006kz}
 \begin{align}
  f_{\alpha R[MN} f_{\beta P]Q}{}^R = 0 \,, \quad
  \epsilon^{\alpha \beta} f_{\alpha MNR} f_{\beta PQ}{}^R = 0 \,.
  \label{QC}
 \end{align}
The first of these should be thought of as the Jacobi identity
leading to closure of the gauge algebra. The other imposes the
orthogonality of charges, i.e.~ensures that one is not using both
the electric and magnetic part of a vector for a gauging, but only a
linear combination.

Note that the commutation relation \eqref{comm} in fact is not
manifestly anti-symmetric on the right-hand side. This is related to
the fact that the 24 generators $X^{M \alpha}$  do not furnish a
basis, as there are only twelve physical gauge vectors and hence the
total gauge algebra can at most be twelve-dimensional. For that
reason there have to be linear relations between the different
generators. These are
 \begin{align}
  \epsilon_{\alpha \beta} f^{\alpha}{}_{MNR} X^{R \beta} = 0 \,.
 \end{align}
Taken in the adjoint representation this is exactly the second
condition of \eqref{QC}. Due to this condition, the right-hand side
of \eqref{comm} is in fact anti-symmetric in the interchange of the
two pairs of indices, as is clear from the left-hand side.

\section{(Non-)geometric flux compactifications}

Now let us see what gauge algebras can be induced by flux
compactifications. The starting point in this discussion are the
results of Kaloper and Myers \cite{KM}. They found that the
dimensional reduction of type I supergravity to four dimensions
leads to a non-Abelian gauge algebra if one includes fluxes. In
particular, they derived the four-dimensional effect of the
following fluxes for the ten-dimensional field content consisting of
the metric, a two-form and a dilaton\footnote{We will only include
fluxes for the metric and the two-form. There is a similar
possibility for the dilaton, which we will not consider, that leads
to gauging with non-vanishing $\xi_{M \alpha}$
\cite{Derendinger:2007xp}. In this paper also the first line of the identification \eqref{STW-ident} was made. Moreover, we will not consider
trombone gaugings of the type introduced in \cite{Diffon} for the
maximal theory.}.

When reducing the metric from ten to four dimensions, one can
generalise ordinary dimensional reduction by replacing the torus
with a group manifold \cite{SS}. A group manifold is specified by
structure constants $\omega_{mn}{}^p$, where the indices run over
the dimension of the group manifold. The four-dimensional effect of
such so-called geometric fluxes is to convert the gauge group
$U(1)^6$, that corresponds to general coordinate transformations on
the torus, to a non-Abelian group with commutation relations
 \begin{align}
  [ Z_m, Z_n ] = \omega_{mn}{}^p Z_p \,,
 \label{KKgenerators}
 \end{align}
where $Z_m$ is the generator corresponding to the internal
coordinate transformation $\delta x^m = \lambda^m$.

Due to the presence of the two-form gauge potential in the
ten-dimensional theory, the four-dimensional gauge algebra is
actually larger. In particular, there is an additional $U(1)^6$
corresponding to internal gauge transformations of the form $\delta
B_{mn} = \partial_{[m} \lambda_{n]}$. We will denote these
generators by $X^p$. These commute amongst themselves, but form a
representation of the group spanned by \eqref{KKgenerators}.
Furthermore, one can introduce gauge fluxes $H_{mnp}$ for this
potential. The total algebra spanned by the six Kaluza-Klein and six
gauge generators reads \cite{KM}
  \begin{align}
   [ Z_m, Z_n] & = \omega_{mn}{}^p Z_p + H_{mnp} X^p \,, \notag \\
   [ Z_m, X^n] & = - \omega_{mp}{}^n X^p \,, \notag \\
   [ X^m ,X^n] & = 0 \,.
  \label{KM-algebra}
  \end{align}
Note that the resulting algebra is purely electric. Furthermore, the
gauge generators span an ideal of the algebra, and hence the full
algebra is non-semi-simple.

In order to make contact with the $SO(6,6)$ notation of $\cN = 4$
supergravity, one needs to split up the $SO(6,6)$ index ${}^M =
({}_m, {}^m)$. The twelve doublets of generators then split up
according to $X^{M \alpha} = (Z_m{}^\alpha, X^{m \alpha})$. The
identification between the embedding tensor and the fluxes is then
apparent:
 \begin{align}
  f_{+mnp} = H_{mnp}  \,, \quad f_{+ mn}{}^p = \omega_{mn}{}^p \,, \label{KM-fluxes}
 \end{align}
while the magnetic components vanish.

A natural question is how to generalise this to the case where one
includes, in addition to gauge and geometric flux, also the types of
non-geometric fluxes introduced by \cite{STW}. If one assumes that
$H$ and $\omega$ are both NS-NS, these will transform into each
other under T-duality. Furthermore, these will transform into the
non-geometric NS-NS fluxes $Q$ and $R$ under T-duality. The action
of T-duality on NS-NS fluxes is to raise and lower the indices of
the different types of fluxes:
 \begin{align}
   H_{mnp} \leftrightarrow \omega_{mn}{}^p \leftrightarrow Q_m{}^{np} \leftrightarrow R^{mnp} \,.
 \end{align}
From this, one can derive what the generalisation of the algebra
\eqref{KM-algebra} is. It can be seen that this reads as \cite{STW}
 \begin{align}
  [ Z_m, Z_n] & = \omega_{mn}{}^p Z_p + H_{mnp} X^p \,, \notag \\
  [ Z_m, X^n] & = - \omega_{mp}{}^n X^p + Q_m{}^{np} Z_p \,, \notag \\
  [ X^m, X^n] & = Q_p{}^{mn} X^p + R^{mnp} Z_p \,. \label{STW-algebra}
 \end{align}
Note that this algebra, with all types of NS-NS fluxes, is still purely electric.

Subsequently one could reason that in the IIB duality frame with
O3-planes one needs to mod out with the $\mathbb{Z}_2$ symmetry
$(-)^{F_L} \Omega I_{4 \cdots 9}$. Under this symmetry, the only
allowed fluxes are $H$ and $Q$. Therefore the algebra for these
fluxes reads
 \begin{align}
  [ Z_m, Z_n] & = H_{mnp} X^p \,, \notag \\
  [ Z_m, X^n] & = Q_m{}^{np} Z_p \,, \notag \\
  [ X^m, X^n] & = Q_p{}^{mn} X^p \,.
 \label{STW-algebra2}
 \end{align}
The relation between the embedding tensor and the fluxes can be
easily read off from this algebra. Before we give it, let us
introduce a slight generalisation by including S-duality related
fluxes as well. For the two-form gauge potentials this is very
natural, as we know that these form a doublet $(H,F)$ under
S-duality. Similarly, it has been conjectured that there is a
doublet of non-geometric fluxes $(Q,P)$ as well
\cite{Aldazabal:2006up}. Including the two doublets of gauge and
non-geometric fluxes, the relation to the embedding tensor is
 \begin{alignat}{2}
  f_{+mnp} & = H_{mnp}  \,, \quad & f_{+ m}{}^{np} & = Q_{m}{}^{np} \,, \notag \\
  f_{-mnp} & = F_{mnp}  \,, \quad & f_{- m}{}^{np} & = P_{m}{}^{np} \,. \label{STW-ident}
 \end{alignat}
The full algebra, including the commutation relations between
electric and magnetic generators, then follows trivially from
\eqref{comm}. Similarly, one can deduce the full set of constraints
on the fluxes from \eqref{QC}.

Note that the algebra \eqref{STW-algebra2} in general does not have
any non-trivial ideals, and hence is not necessarily
non-semi-simple. This form of the algebra has been used in e.g.
\cite{deCarlos1} in their classification of the possible solutions
of the corresponding Jacobi identities. Indeed, they encountered
simple and semi-simple possibilities. This poses a clear puzzle: we
claim to have performed a number of dualities, under which the
effective description should transform covariantly, and nevertheless
the algebra \eqref{KM-algebra} of the starting point clearly differs
from \eqref{STW-algebra2}. Indeed, one is necessarily
non-semi-simple while the other is not. What has happened? In our
opinion, the confusion stems from the identification of the starting
point.

The starting point of Kaloper and Myers corresponds to the heterotic
string, and therefore contains an NS-NS two-form gauge potential.
However, in order to make contact with type II string theories with
orientifold planes, e.g.~the preferred duality frame of type IIB
with O3-planes, one should first perform an S-duality. This takes
one to type I string theory, or equivalently type IIB with
O9-planes. In this case the two-form is not NS-NS but rather R-R,
which will be a crucial distinction when applying T-duality. As
mentioned before, in the NS-NS sector T-duality raises and lowers
indices. In contrast, in the R-R sector the effect of T-duality is
to create or annihilate indices:
\begin{align}
 T_p: \qquad
\begin{cases}
    F_{m_1 \cdots m_n} \rightarrow F_{m_1 \cdots m_n p} \,, \\
 F_{m_1 \cdots m_n p} \rightarrow F_{m_1 \cdots m_n} \,,
\end{cases}
\end{align}
In other words, a gauge potential remains a gauge potential but its
rank changes.

The correct starting point for our purpose is
  \begin{align}
   [ Z_m, Z_n] & = \omega_{mn}{}^p Z_p + F_{mnp} X^p \,, \notag \\
   [ Z_m, X^n] & = - \omega_{mp}{}^n X^p \,, \notag \\
   [ X^m ,X^n] & = 0 \,,
  \label{KM-algebra2}
  \end{align}
where $F_{mnp}$ is the R-R three-form flux. Upon a six-tuple
T-duality to go to the type IIB duality frame with O3-planes, this
transforms into
 \begin{align}
  [ Z_m, Z_n] & = 0  \,, \notag \\
  [ Z_m, X^n] & = Q_{m}{}^{np} Z_p \,, \notag \\
  [ X^m ,X^n] & = Q_p{}^{mn} X^p +  {\tilde F}^{mnp} Z_p \,.
 \label{ACR-algebra1}
 \end{align}
where ${\tilde F}^{mnp} = \tfrac16 \epsilon^{mnpqrs} F_{qrs}$. This
fixes the complete electric part of the gauge algebra. The remaining
part follows straightforwardly once one has made the identification
between the embedding tensor and the fluxes. Again we will give an
S-duality covariant set of fluxes, including the gauge doublet
$(F,H)$ and the non-geometric doublet $(Q,P)$. With the algebra
\eqref{ACR-algebra1} this identification reads
 \begin{alignat}{2}
  f_{+}{}^{mnp} & = \tilde{F}^{mnp}  \,, \quad & f_{+ m}{}^{np} & = Q_{m}{}^{np} \,, \notag \\
  f_{-}{}^{mnp} & = \tilde{H}^{mnp}  \,, \quad & f_{- m}{}^{np} & = P_{m}{}^{np} \,. \label{ACR-ident}
 \end{alignat}
The full algebra and corresponding quadratic constraints then follow
from \eqref{comm} and \eqref{QC}. The latter read
 \begin{align}
  & Q_r{}^{[mn}Q_q{}^{p]r} =   P_r{}^{[mn}P_q{}^{p]r}= 0 \,, \notag \\
  & P_r{}^{[mn}Q_q{}^{p]r} = Q_r{}^{mn}P_q{}^{pr} - P_r{}^{mn}Q_q{}^{pr} = 0  \,, \label{QC1}
 \end{align}
involving only non-geometric flux, and
 \begin{align}
  & \tilde{F}^{r[mn}Q_r{}^{pq]} =
  \tilde{H}^{r[mn}P_r{}^{pq]} = 0 \,, \notag \\
 &  \tilde{F}^{r[mn}P_r{}^{p]q}+ Q_r{}^{[mn}\tilde{H}^{p]qr}  = 0  \,, \label{QC2}
\end{align}
involving gauge fluxes as well. The fully anti-symmetric parts of
the latter set of equations imply the absence of any 7-branes; these
would break supersymmetry further to $\cN = 1$. The same form  of
the algebra and quadratic constraints was recently derived in the
beautiful work\footnote{Due to different conventions regarding the
$SO(6,6)$ and $SL(6)$ indices, our form of the identification
\eqref{ACR-ident} does not involve any non-trivial metrics, as in
\cite{ACR}. Moreover, the quadratic constraints given in \cite{ACR}
are not all linearly independent, and hence can be written in a more
economic way.} \cite{ACR} from a different starting point.

Note the differences between the two algebras\footnote{Most of the
literature that uses \eqref{STW-algebra} takes place in an $\cN = 1$
context, where the scalar potential is not given in terms of
structure constants but rather a superpotential. Therefore our
argument does not affect any of the results on $\cN =1$ moduli
stabilisation etc.} \eqref{STW-algebra} and \eqref{ACR-algebra1}.
First of all, NS-NS fluxes induce a purely electric gauging in the
former algebra \cite{Derendinger:2007xp}, while in the latter this
involves magnetic generators as well. Moreover, the former can
describe a (semi-)simple algebra
(see e.g.~\cite{Prezas1,Prezas2,deCarlos1}), while the latter
is always non-semi-simple algebra, as it should. This crucial
difference between the two stems from the appearance of the Hodge
dualised three-form $\tilde F$, instead of the three-form itself, in
\eqref{ACR-algebra1}. This qualitative difference can be traced back
to the different behaviour of NS-NS and R-R gauge potentials under
T-duality.

Finally, the quadratic constraints \eqref{QC2} are in general
different for the two algebras. For instance, it can be seen from
the $SL(2)$ scaling weight that the last equation of \eqref{QC2}
could never arise from \eqref{STW-ident}. However, in the truncation
where one of the two non-geometric fluxes vanishes, e.g. $P=0$, the
quadratic constraints bilinear in the NS-NS fluxes are in fact
identical (provided $Q_m{}^{mn} = 0$). There is still a difference
in the constraints bilinear in $Q$ and $F$: these are much stronger
for the first identification \eqref{STW-ident} than those given in
\eqref{QC2}.

\section{What about De Sitter?}

All the gaugings that are known to give rise to De Sitter solutions
in $\cN = 4$ gauged supergravity \cite{deRoo:2002jf, deRoo:2003rm}
are of the form
 \begin{align}
  G = G_1 \times G_2 \times \cdots \,, \label{gaugegroup}
 \end{align}
i.e.~a direct product of a number of gauge factors. This is a
solution to the quadratic constraints \eqref{QC} once the Jacobi
identities are separately satisfied in the different factors.
Moreover, in order to have a De Sitter solution, the gauge group
must contain electric and magnetic factors. Finally, the gauge
factors have to be specific (semi-)simple groups. In particular, we
will focus on the case of two gauge factors. Each factor is  of the
form $SO(p,q)$ with $p+q=4$ and embedded in an $SO(3,3)$ factor. A
number of examples of such gaugings with De Sitter solutions was
discussed in \cite{deRoo:2002jf, deRoo:2003rm}. Moreover, it was
shown in \cite{deRoo:2006ms} that the contracted versions
$CSO(p,q,r)$ with $p+q+r = 4$ of such gauge groups do not lead to
any solutions with a positive scalar potential.  In this section we
will assess to what extend one can obtain such gaugings from the
flux compactifications considered earlier.

The direct product structure \eqref{gaugegroup} leads us to split
$SO(6,6)$ into two $SO(3,3)$ factors in which to embed $G_{1}$ and
$G_{2}$ respectively. Without loss of generality, we will take the
first to be electric and lie in the directions $\{1,2,3,\bar 1 ,
\bar 2 , \bar 3 \}$, while the second is taken magnetic and lies in
the complementary directions. We will discuss the embedding of the
first factor in some detail; the discussion for the second factor is
completely analogous. However, before we discuss $SO(4)$ embeddings
in $SO(3,3) \simeq SL(4)$, we first generalise this to arbitrary
$N$.

In general, the embedding of $SO(N)$ and its analytic continuations
into $SL(N)$ can be written in terms of the following generators in
the fundamental representation
\begin{align}
\left(T_{ij}\right)^{k}_{\phantom{k}l}= 4 \delta^{k}_{\phantom{k}[i} M_{j]l}\,,
\end{align}
in terms of a symmetric matrix $M$, that can always be diagonalised
by a convenient choice of basis. It is in fact given by the identity
in the case of $SO(N)$. These generators labelled by antisymmetric
pairs of indices satisfy the following commutation relations
\begin{align}
\left[T_{ij},T_{kl}\right]=f_{ij,kl}^{\phantom{ij,kl}mn} T_{mn}\,,\quad
f_{ij,kl}^{\phantom{ij,kl}mn}= 8 \delta^{[m}_{\phantom{k}[i} M_{j][k}\delta^{n]}_{\phantom{k}l]}\,.
\end{align}
Analytic continuations of $SO(N)$  correspond to a number of minus
signs in the $M$-matrix. Contractions thereof, denoted by
$CSO(p,q,r)$ with $p+q+r=N$ (see e.g.~\cite{deRoo:2006ms}), can be
understood in  this notation by replacing $r$ non-zero diagonal
entries of $M$ with zero entries.

However, the most general form of $CSO(p,q,r)$ structure constants
for the special case of $N = 4$ is given in terms of two symmetric
matrices rather than one \cite{Rosseel}, which we will denote by $M$
and $\tilde{M}$. The generators are then given by
\begin{align}
\left(T_{ij}\right)^{k}_{\phantom{k}l}= 4 \delta^{k}_{\phantom{k}[i} M_{j]l} - 2 \varepsilon_{ijml}\tilde{M}^{mk}\,,
\end{align}
giving rise to the following general expression of the structure
constants
\begin{align}
f_{ij,kl}^{\phantom{ij,kl}mn}= 8 \delta^{[m}_{\phantom{k}[i} M_{j][k}\delta^{n]}_{\phantom{k}l]} - \varepsilon_{iji'j'}\varepsilon_{klk'l'}\varepsilon^{mni'l'} \tilde{M}^{j'k'} \,.
\end{align}
With such a form we need some extra consistency constraints in terms
of $M$ and $\tilde{M}$, coming from imposing the Jacobi
identities. These translate into
\begin{align}
M_{ij}\tilde{M}^{jk}-\tfrac{1}{4}\delta_i^{\phantom{a}k}M_{jl}\tilde{M}^{jl}=0\,.
\end{align}
If one still diagonalises $M$ by a convenient basis choice, the
Jacobi identity imply $\tilde{M}$ to be diagonal as well. In this
case the constraints reduce to
\begin{align}
M_{11}\tilde{M}^{11}=M_{22}\tilde{M}^{22}=M_{33}\tilde{M}^{33}=M_{44}\tilde{M}^{44}\,.
\end{align}

Let us now connect the adjoint representation in terms of $SL(4)$
indices to fundamental $SO(3,3)$ indices. This relation is given by
 \begin{align}
  \{ 1,2,3,\bar 1, \bar 2, \bar 3 \} \simeq \{ 12,13,14,43,24,32 \} \,.
 \end{align}
This leads to the following identification between the diagonal
components of the two matrices $M$ and $\tilde M$, and the
components of the embedding tensor $f_{\alpha MNP}$ in the first
$SO(3,3)$ factor:
 \begin{align}
  M & = \textrm{diag}(f_{+123}, f_{+1\bar2 \bar 3}, f_{+ \bar 1 2 \bar 3}, f_{+\bar 1 \bar 2 3}) \,, \notag \\
 \tilde{M} & =  \textrm{diag}(f_{+ \bar 1 \bar 2 \bar 3}, f_{+ \bar 1 2   3}, f_{+   1  \bar 2   3}, f_{+  1   2  \bar 3}) \,.
 \end{align}
Other components of the embedding tensor in this $SO(3,3)$ factor,
such as $f_{+ 1 \bar 1 2}$ and $f_{+ 1 \bar 1 \bar 2}$, correspond to off-diagonal components of
$M$ and $\tilde M$ and hence have been set equal to zero.

We have discussed in the previous sections how the embedding tensor
can be sourced by different fluxes. In particular, we have discussed
the two identifications \eqref{STW-ident} and \eqref{ACR-ident}. It
will be illuminating to illustrate the different consequences of the
two identifications in this context. Using the first identification,
the matrices are given by
 \begin{align}
  M & = \textrm{diag} (H_{123} , Q_{1}{}^{23} , Q_{2}{}^{31}, Q_{3}{}^{12}) \,,  \notag \\
 \tilde{M}  & =  \textrm{diag}(0,0,0,0) \,.
 \end{align}
In this case it would therefore be possible to use the different
fluxes to generate a simple gauge factor. Given that the discussion
in the second, magnetic factor is completely analogous, one could
e.g.~generate an $SO(4)_{\rm el} \times SO(4)_{\rm magn}$ gauge
group, which certainly leads to De Sitter solutions. However, we
have argued that this is not the correct identification; instead,
one should use \eqref{ACR-ident}. In this case, the matrices read
 \begin{align}
  M & = \textrm{diag} (0 , Q_{1}{}^{23} , Q_{2}{}^{31}, Q_{3}{}^{12}) \,,  \notag \\
\tilde{M}  & =  \textrm{diag}(F_{456},0,0,0) \,.
 \end{align}
The crucial point is that in this case the gauge flux does not enter
in the $M$ matrix to make it non-singular; instead, it enters in the
other matrix $\tilde M$. These singular matrices only lead to
non-semi-simple gauge groups. In particular, the matrix $M$ gives
rise to $ISO(3)$ and analytic continuations and contractions
thereof. Provided the three components $Q_i{}^{jk}$ are non-zero,
the additional parameter $F_{456}$ does not modify the gauge group,
but only describes different embeddings of it in $SO(3,3)$. Three of
these are inequivalent, corresponding to $F_{456}$ being positive,
zero or negative. Exactly the same embeddings of $ISO(3)$ and
$ISO(2,1)$ were considered in\footnote{The relation to the notation
of  \cite{deRoo:2006ms} is $\lambda^2 = (1-F_{456})/(1+F_{456})$.}
\cite{deRoo:2006ms}, where it was found that such gauge groups do
not give rise to scalar potentials with positive extrema.

Indeed, one can infer from the same reasoning that none of the gauge
groups discussed in \cite{deRoo:2002jf, deRoo:2003rm} follows from a
flux compactification with the identification \eqref{ACR-ident}. The
simple bottom line is that all the gauge groups necessarily consist
of (semi-)simple gauge factors, while one can only get
non-semi-simple factors from flux compactifications.

\section{Conclusions}

One of the main points of this letter is to point out the gauge
algebra \eqref{ACR-algebra1} that arises from (non-)geometric flux
compactifications to $D= \cN = 4$. In contrast to
\eqref{STW-algebra}, this algebra is always non-semi-simple due to
the presence of an Abelian ideal spanned by the generators $Z_m$. As
a consequence, it is impossible to build any of the gauge groups
consisting of simple factors that are known to give rise to De
Sitter solutions \cite{deRoo:2002jf, deRoo:2003rm}.

There is a number of directions in which to extend this work.
Amongst them are generalisations of the flux compactifications and
gauge algebras discussed in section 3. For instance, one could
include the world-volume excitations of D3-branes to change the
number of $\cN = 4$ vector multiplets. Similarly, one could consider
the truncation to $\cN = 1$ supergravity by including O7-planes and
D7-branes. Some aspects of these extensions can be found in
\cite{ACR}. Finally, one could consider going beyond the type of
flux compactifications discussed here\footnote{One could e.g.~investigate the possibility of O3-planes on doubled geometry \cite{Hull1,Hull2,Spanjaard} with non-geometric fluxes. In contrast to the heterotic duality frame, where non-geometric fluxes and doubled geometry give rise to the same generalisation of \cite{KM}, in the O3 duality frame doubled geometry could lead to a doubling of the flux components in \eqref{ACR-ident}. In such a ``doubled non-geometric'' set-up all components of the $\cN = 4$ embedding tensor could be turned on.} to account for the missing
components of the embedding tensor in \eqref{ACR-ident}, and in this
way build up (semi-)simple gauge algebras.

As for the possibilities of De Sitter, again a number of
generalisations are possible. In \cite{deRoo:2002jf, deRoo:2003rm}
an analysis was made which gauge groups lead to a positive
cosmological constant in the origin. Naturally, this could be
extended to a larger portion of the moduli space. Indeed, such an
analysis was performed in the very recent work \cite{deCarlos1,
deCarlos2} for $\cN = 1$ flux compactifications with $P=0$. In a
clever way all possible Minkowski vacua were determined, and a band
of De Sitter vacua was found closeby (in moduli and parameter
space). It can be seen that one of their cases\footnote{There is in
fact a close link between another of their cases, where the $Q$-flux
spans an $ISO(3)$ algebra, and the unstable De Sitter solution found
in \cite{Caviezel:2008tf} from a IIA flux compactification on an
$SU(2) \times SU(2)$ group manifold. We have checked explicitly that
the fluxes given in the latter paper correspond to the angles
$\theta_\xi=0.38 \pi$ and $\theta_\epsilon = 1.47 \pi$ in the
notation of \cite{deCarlos2}. Therefore the single example of an
unstable De Sitter solution can be understood to lie in a narrow
band of such solutions that borders the line of unstable Minkowski
solutions.}, where the $Q$-flux spans an $SO(3,1)$ algebra, allows
for an interpretation in terms of $\cN = 4$ as well; in this case,
all quadratic constraints \eqref{QC1} and \eqref{QC2} can be
fulfilled. Therefore it is possible to obtain De Sitter solutions
from $\cN = 4$ non-geometric compactifications. A natural question
concerns the gauge algebra in this case; in other words, given the
fluxes, what algebra does \eqref{ACR-algebra1} correspond to? It
appears that it is no longer of the direct product form
\eqref{gaugegroup} but rather a semi-direct product, where e.g.~the
electric part of the gauge group has a non-trivial action on the
magnetic part. We leave this question for future investigation.

\section*{Acknowledgements}

We would like to thank Gerardo Aldazabal, Pablo C\'{a}mara, Beatriz
de Carlos, Adolfo Guarino, Jes\'{u}s M.~Moreno and Alejandro Rosabal
for useful correspondence and furthermore Adolfo Guarino for
stimulating discussions. RL and DR would like to thank each other's
institutes for warm hospitality. The work of GD and DR is supported
by a VIDI grant from the Netherlands Organisation for Scientific
Research (NWO). The work of RL was partially supported by Mexico's
National Council of Science and Technology under grant
CONACyT-SEP-2004-C01-47597 and by the Erasmus Mundus ECW
Mexico-Program.

\providecommand{\href}[2]{#2}\begingroup\raggedright\endgroup


\begin{thebibliography}{10}

\bibitem{Grana}
M.~Grana,  {\em {Flux compactifications in string theory: A comprehensive
  review}}, Phys. Rept. {\bf 423} (2006) 91--158
[\href{http://www.arXiv.org/abs/hep-th/0509003}{{hep-th/0509003}}].

\bibitem{Douglas}
M.~R. Douglas and S.~Kachru,  {\em {Flux compactification}}, Rev. Mod. Phys.
  {\bf 79} (2007) 733--796
[\href{http://www.arXiv.org/abs/hep-th/0610102}{{hep-th/0610102}}].

\bibitem{Blumenhagen}
R.~Blumenhagen, B.~Kors, D.~Lust and S.~Stieberger,  {\em {Four-dimensional
  String Compactifications with D-Branes, Orientifolds and Fluxes}}, Phys.
  Rept. {\bf 445} (2007) 1--193
[\href{http://www.arXiv.org/abs/hep-th/0610327}{{hep-th/0610327}}].

\bibitem{Hertzberg:2007wc}
M.~P. Hertzberg, S.~Kachru, W.~Taylor and M.~Tegmark,  {\em {Inflationary
  Constraints on Type IIA String Theory}}, JHEP {\bf 12} (2007) 095
[\href{http://www.arXiv.org/abs/0711.2512}{{0711.2512}}].

\bibitem{Silverstein}
E.~Silverstein,  {\em {Simple de Sitter Solutions}}, Phys. Rev. {\bf D77}
  (2008) 106006
[\href{http://www.arXiv.org/abs/0712.1196}{{0712.1196}}].

\bibitem{Haque:2008jz}
S.~S. Haque, G.~Shiu, B.~Underwood and T.~Van~Riet,  {\em {Minimal simple de
  Sitter solutions}}, Phys. Rev. {\bf D79} (2009) 086005
[\href{http://www.arXiv.org/abs/0810.5328}{{0810.5328}}].

\bibitem{Caviezel:2008tf}
C.~Caviezel {\em et al.},  {\em {On the Cosmology of Type IIA Compactifications
  on SU(3)- structure Manifolds}}, JHEP {\bf 04} (2009) 010
[\href{http://www.arXiv.org/abs/0812.3551}{{0812.3551}}].

\bibitem{Flauger:2008ad}
R.~Flauger, S.~Paban, D.~Robbins and T.~Wrase,  {\em {Searching for slow-roll
  moduli inflation in massive type IIA supergravity with metric fluxes}}, Phys.
  Rev. {\bf D79} (2009) 086011
[\href{http://www.arXiv.org/abs/0812.3886}{{0812.3886}}].

\bibitem{Danielsson}
U.~H. Danielsson, S.~S. Haque, G.~Shiu and T.~Van~Riet,  {\em {Towards
  Classical de Sitter Solutions in String Theory}}, JHEP {\bf 09} (2009) 114
[\href{http://www.arXiv.org/abs/0907.2041}{{0907.2041}}].

\bibitem{Zagermann}
C.~Caviezel, T.~Wrase and M.~Zagermann,  {\em {Moduli Stabilization and
  Cosmology of Type IIB on SU(2)- Structure Orientifolds}},
\href{http://www.arXiv.org/abs/0912.3287}{{0912.3287}}.

\bibitem{Roest:2009dq}
D.~Roest,  {\em {Gaugings at angles from orientifold reductions}}, Class.
  Quant. Grav. {\bf 26} (2009) 135009
[\href{http://www.arXiv.org/abs/0902.0479}{{0902.0479}}].

\bibitem{Dall'Agata:2009gv}
G.~Dall'Agata, G.~Villadoro and F.~Zwirner,  {\em {Type-IIA flux
  compactifications and N=4 gauged supergravities}}, JHEP {\bf 08} (2009) 018
[\href{http://www.arXiv.org/abs/0906.0370}{{0906.0370}}].

\bibitem{STW}
J.~Shelton, W.~Taylor and B.~Wecht,  {\em {Nongeometric Flux
  Compactifications}}, JHEP {\bf 10} (2005) 085
[\href{http://www.arXiv.org/abs/hep-th/0508133}{{hep-th/0508133}}].

\bibitem{Aldazabal:2006up}
G.~Aldazabal, P.~G. Camara, A.~Font and L.~E. Ibanez,  {\em {More dual fluxes
  and moduli fixing}}, JHEP {\bf 05} (2006) 070
[\href{http://www.arXiv.org/abs/hep-th/0602089}{{hep-th/0602089}}].

\bibitem{Hull1}
C.~M. Hull,  {\em {Doubled geometry and T-folds}}, JHEP {\bf 07} (2007) 080
[\href{http://www.arXiv.org/abs/hep-th/0605149}{{hep-th/0605149}}].

\bibitem{Hull2}
C.~M. Hull and R.~A. Reid-Edwards,  {\em {Gauge Symmetry, T-Duality and Doubled
  Geometry}}, JHEP {\bf 08} (2008) 043
[\href{http://www.arXiv.org/abs/0711.4818}{{0711.4818}}].

\bibitem{Grana:2006hr}
M.~Grana, J.~Louis and D.~Waldram,  {\em {SU(3) x SU(3) compactification and
  mirror duals of magnetic fluxes}}, JHEP {\bf 04} (2007) 101
[\href{http://www.arXiv.org/abs/hep-th/0612237}{{hep-th/0612237}}].

\bibitem{deCarlos1}
B.~de~Carlos, A.~Guarino and J.~M. Moreno,  {\em {Flux moduli stabilisation,
  Supergravity algebras and no-go theorems}}, JHEP {\bf 01} (2010) 012
[\href{http://www.arXiv.org/abs/0907.5580}{{0907.5580}}].

\bibitem{deCarlos2}
B.~de~Carlos, A.~Guarino and J.~M. Moreno,  {\em {Complete classification of
  Minkowski vacua in generalised flux models}},
\href{http://www.arXiv.org/abs/0911.2876}{{0911.2876}}.

\bibitem{deRoo:2002jf}
M.~de~Roo, D.~B. Westra and S.~Panda,  {\em {De Sitter solutions in N = 4
  matter coupled supergravity}}, JHEP {\bf 02} (2003) 003
[\href{http://www.arXiv.org/abs/hep-th/0212216}{{hep-th/0212216}}].

\bibitem{deRoo:2003rm}
M.~de~Roo, D.~B. Westra, S.~Panda and M.~Trigiante,  {\em {Potential and
  mass-matrix in gauged N = 4 supergravity}}, JHEP {\bf 11} (2003) 022
[\href{http://www.arXiv.org/abs/hep-th/0310187}{{hep-th/0310187}}].

\bibitem{Samtleben}
H.~Samtleben,  {\em {Lectures on Gauged Supergravity and Flux
  Compactifications}}, Class. Quant. Grav. {\bf 25} (2008) 214002
[\href{http://www.arXiv.org/abs/0808.4076}{{0808.4076}}].

\bibitem{Schon:2006kz}
J.~Schon and M.~Weidner,  {\em {Gauged N = 4 supergravities}}, JHEP {\bf 05}
  (2006) 034
[\href{http://www.arXiv.org/abs/hep-th/0602024}{{hep-th/0602024}}].

\bibitem{KM}
N.~Kaloper and R.~C. Myers,  {\em {The O(dd) story of massive supergravity}},
  JHEP {\bf 05} (1999) 010
[\href{http://www.arXiv.org/abs/hep-th/9901045}{{hep-th/9901045}}].

\bibitem{Derendinger:2007xp}
J.-P. Derendinger, P.~M. Petropoulos and N.~Prezas,  {\em {Axionic symmetry
  gaugings in N = 4 supergravities and their higher-dimensional origin}}, Nucl.
  Phys. {\bf B785} (2007) 115--134
[\href{http://www.arXiv.org/abs/0705.0008}{{0705.0008}}].

\bibitem{Diffon}
A.~Le~Diffon and H.~Samtleben,  {\em {Supergravities without an Action: Gauging
  the Trombone}}, Nucl. Phys. {\bf B811} (2009) 1--35
[\href{http://www.arXiv.org/abs/0809.5180}{{0809.5180}}].

\bibitem{SS}
J.~Scherk and J.~H. Schwarz,  {\em {How to Get Masses from Extra Dimensions}},
  Nucl. Phys. {\bf B153} (1979)
61--88.

\bibitem{ACR}
G.~Aldazabal, P.~G. Camara and J.~A. Rosabal,  {\em {Flux algebra, Bianchi
  identities and Freed-Witten anomalies in F-theory compactifications}}, Nucl.
  Phys. {\bf B814} (2009) 21--52
[\href{http://www.arXiv.org/abs/0811.2900}{{0811.2900}}].

\bibitem{Prezas1}
G.~Dall'Agata and N.~Prezas,  {\em {Worldsheet theories for non-geometric
  string backgrounds}}, JHEP {\bf 08} (2008) 088
[\href{http://www.arXiv.org/abs/0806.2003}{{0806.2003}}].

\bibitem{Prezas2}
S.~D. Avramis, J.-P. Derendinger and N.~Prezas,  {\em {Conformal chiral boson
  models on twisted doubled tori and non-geometric string vacua}}, Nucl. Phys.
  {\bf B827} (2010) 281--310
[\href{http://www.arXiv.org/abs/0910.0431}{{0910.0431}}].

\bibitem{deRoo:2006ms}
M.~de~Roo, D.~B. Westra and S.~Panda,  {\em {Gauging CSO groups in N = 4
  supergravity}}, JHEP {\bf 09} (2006) 011
[\href{http://www.arXiv.org/abs/hep-th/0606282}{{hep-th/0606282}}].

\bibitem{Rosseel}
D.~Roest and J.~Rosseel,  {\em {De Sitter in Extended Supergravity}}, Phys.
  Lett. {\bf B685} (2010) 201--207
[\href{http://www.arXiv.org/abs/0912.4440}{{0912.4440}}].

\bibitem{Spanjaard}
R.~A. Reid-Edwards and B.~Spanjaard,  {\em {N=4 Gauged Supergravity from
  Duality-Twist Compactifications of String Theory}}, JHEP {\bf 12} (2008) 052
[\href{http://www.arXiv.org/abs/0810.4699}{{0810.4699}}].

\end{thebibliography}
\end{document}